\documentclass[aps,prx,twocolumn,showpacs,groupedaddress,notitlepage,superscriptaddress,floatfix,longbibliography]{revtex4-1}
\usepackage{amsmath,amsthm,amssymb,amsfonts,enumitem,float,graphicx,physics,hyperref}%, xcolor}

\usepackage{color}
\usepackage[dvipsnames]{xcolor}

\newcommand{\beq}{\begin{equation}}
\newcommand{\eeq}{\end{equation}}
\newcommand{\bes} {\begin{subequations}}
\newcommand{\ees} {\end{subequations}}
\newcommand{\pvec}[1]{\vec{#1}\mkern2mu\vphantom{#1}}

\hypersetup{
    colorlinks=true, 
    linkcolor=cyan,
    citecolor=magenta, 
    filecolor=magenta, 
    urlcolor=cyan,
    runcolor=cyan
}

\begin{document}

\title{Quantum adiabatic machine learning with zooming}

\author{Alexander Zlokapa}
\affiliation{Division of Physics, Mathematics \& Astronomy, Alliance for Quantum Technologies, California Institute of Technology, Pasadena, CA 91125, USA}

\author{Alex Mott}
\affiliation{DeepMind Technologies, London, UK}

\author{Joshua Job}
\affiliation{Lockheed Martin Advanced Technology Center, Sunnyvale, CA 94089, USA}

\author{Jean-Roch Vlimant}
\affiliation{Division of Physics, Mathematics \& Astronomy, Alliance for Quantum Technologies, California Institute of Technology, Pasadena, CA 91125, USA}

\author{Daniel Lidar}
\affiliation{Departments of Electrical and Computer Engineering, Chemistry, and Physics \& Astronomy, and Center for Quantum  Information Science \& Technology, University of Southern California, Los Angeles, CA 90089, USA}

\author{Maria Spiropulu}
\affiliation{Division of Physics, Mathematics \& Astronomy, Alliance for Quantum Technologies, California Institute of Technology, Pasadena, CA 91125, USA}

%\keywords{Keyword1, Keyword2, Keyword3}

\begin{abstract}
Recent work has shown that  quantum annealing for machine learning, referred to as QAML, can perform comparably to state-of-the-art machine learning methods with a specific application to Higgs boson classification. We propose QAML-Z, a novel algorithm  that iteratively zooms in on a region of the energy surface by mapping the problem to a continuous space and sequentially applying quantum annealing to an augmented set of weak classifiers. Results on a programmable quantum annealer show that QAML-Z matches classical deep neural network performance at small training set sizes and reduces the performance margin between QAML and classical deep neural networks by almost 50\% at large training set sizes, as measured by area under the ROC curve. The significant improvement of quantum annealing algorithms for machine learning and the use of a discrete quantum algorithm on a continuous optimization problem both opens a new class of problems that can be solved by quantum annealers and suggests the approach in performance of near-term quantum machine learning towards classical benchmarks.
\end{abstract}

\maketitle

\section{Introduction}%
Machine learning has gained an increasingly important role in scientific discovery across chemistry, biology, environmental science, and physics~\cite{leelananda2016computational, tarca2007machine, hsieh2009machine, carrasquilla2017machine, baldi2014searching}, including in the discovery of the Higgs boson~\cite{radovic2018machine}. Various quantum computing algorithms have been proposed for machine learning~\cite{biamonte2017quantum}, including support vector machines, principal component analysis, least-squares fitting, topological analysis, and other optimization problems~\cite{rebentrost2014quantum, lloyd2014quantum, wiebe2012quantum, schuld2016prediction, lloyd2016quantum, rebentrost2019quantum}. Many of these algorithms include strict data assumptions that provide critical caveats regarding sparsity, state preparation, and rank~\cite{aaronson2015read, tang2019quantum}. Moreover, fault-tolerant quantum computing will be required to implement the large quantum circuits necessary for the proposed algorithms, which has not yet been experimentally established at a scale necessary for the implementation of the proposed algorithms. Similarly, quantum random access memory (qRAM) is typically required to store classical data, but engineering challenges persist in developing a sufficiently large qRAM~\cite{arunachalam2015robustness}.

One promising near-term avenue for quantum machine learning is quantum annealing~\cite{kadowaki1998quantum} (for recent reviews see~\cite{Job:2017aa,Hauke:2019aa, job2018theory}) which can, e.g., perform binary classification~\cite{neven2008training,qml}, learn Bayesian network structure~\cite{OGorman:2015qf}, implement quantum Boltzmann machines~\cite{Amin:2016}, and train deep generative models~\cite{vinci2019path}. Quantum annealing is the only current quantum computing paradigm that has resulted in architectures with a large enough number of --- albeit relatively noisy --- qubits~\cite{harris2010experimental, johnson2011quantum, dwave} to address both real-world and fundamental science problems, e.g., in air traffic control~\cite{stollenwerk2019quantum}, computational biology~\cite{perdomo2012finding, li2018quantum, Li:2019aa}, and high energy physics~\cite{nature,zlokapa2019charged,Bapst_2019}. Under the adiabatic theorem of quantum mechanics, quantum annealing evolves from an initial transverse field Hamiltonian to the target problem Hamiltonian, ensuring that the system remains in the ground state if the system is perturbed slowly enough, as given by the energy gap between the ground state and the first excited state~\cite{kato1950adiabatic,Jansen:07,lidar:102106}. The ground state of the problem Hamiltonian is then the solution (as in adiabatic quantum computing~\cite{Farhi:00,albash2018adiabatic}), although thermal excitations 
%and relaxation as well as magnetic coupling of physical qubits to logical qubits 
may move the system out of the ground state~\cite{childs2001robustness, amin2009decoherence, albash2015decoherence, embed1, embed2, klymko_adiabatic_2012, Cai:2014nx},which can be beneficial~\cite{TAQC,dickson2013thermally,Venuti:2017aa,Mishra:2018}.
It is crucial to observe that evidence of a quantum speedup in quantum annealing remains uncertain~\cite{speedup,Albash:2017aa,Mandra:2017ab}, although quantum phenomena have been observed in D-Wave quantum annealers~\cite{lanting2014entanglement,Boixo:2014yu,PhysRevX.6.031015}. While this remains a speculative topic, quantum annealers may exhibit advantages other than a speedup, 
%as result of these phenomena, 
such as sampling from non-equilibrium distributions prepared during the anneal~\cite{Amin:2015qf,King:2018aa,Harris:2018aa}.

Here we propose a novel quantum algorithm inspired by the previous state-of-the-art quantum annealing for machine learning (QAML) algorithm~\cite{nature}, which constructs a single strong classifier from a linear combination of weak classifiers with binary coefficients of 1 or 0. We propose two modifications to QAML — zooming into the energy surface to optimize real-valued coefficients and artificially augmenting the set of weak classifiers to create a stronger ensemble — and implement the proposed algorithm (QAML-Z) on the D-Wave quantum annealer to benchmark the results on a Higgs classification problem, with available source code~\footnote{\label{ref:src}Source code for the QAML-Z algorithm is provided at \url{https://github.com/quantummind/qaml-z/}.} and data~\cite{data}.

\section{QAML-Z Algorithm}%
\subsection{Background: QAML Algorithm}
In the original QAML algorithm, a training set with $S$ examples of labeled data $\{\mathbf{x}_\tau, y_\tau\}$ (where $\mathbf{x}_\tau$ is an input vector and $y_\tau = \pm 1$ is a binary label for signal and background) is optimized with a set of $N$ weak classifiers $c_i$, each of which gives $c_i(\mathbf{x}_\tau) = \pm 1/N$ for a signal or background prediction. Given spins $s_i \in \{0, 1\}$ obtained by transforming up/down spins, let $R(\mathbf{x}_\tau)$ be a strong classifier given by
\begin{equation}
    R(\mathbf{x}_\tau) = \sum_{i=1}^N s_ic_i(\mathbf{x}_\tau) ,
\end{equation}
i.e., an ensemble of the weak classifiers where each weak classifier is either turned on or off (weight 1 or 0). To minimize classification error, we simply minimize the distance between $y$ and $R$:
\bes
\begin{align}
    ||\mathbf{y} - \mathbf{R}||^2 &= \sum_{\tau=1}^S \left|y_\tau - \sum_{i=1}^N s_i c_i(\mathbf{x}_\tau) \right|^2\\
    &= ||\mathbf{y}||^2 - 2\sum_{i=1}^N \sum_{\tau=1}^S s_i c_i(\mathbf{x}_\tau)y_\tau \notag\\
    &\;\;\;\;+ \sum_{i=1}^N\sum_{j=1}^N \sum_{\tau=1}^S s_i c_i(\mathbf{x}_\tau)s_j c_j(\mathbf{x}_\tau) .
\end{align}
\ees
Removing the spin-independent term $||\mathbf{y}||^2$ and the self-spin interactions $c_i^2(\mathbf{x}_\tau)$ to construct a problem suitable for quantum annealing, we rewrite the Hamiltonian as follows (scaling by a factor of 2 for convenience after manipulating indices):
\begin{equation}
    H = \sum_{i=1}^N\sum_{j>i}^N \sum_{\tau=1}^S s_i c_i(\mathbf{x}_\tau)s_j c_j(\mathbf{x}_\tau) - \sum_{i=1}^N \sum_{\tau=1}^S s_i c_i(\mathbf{x}_\tau)y_\tau .
\end{equation}
For convenience, we define the variables:
\begin{align}
    C_{ij} &= \sum_{\tau=1}^S c_i(\mathbf{x}_\tau)c_j(\mathbf{x}_\tau),\\
    C_i &= \sum_{\tau=1}^S c_i(\mathbf{x}_\tau)y_\tau .
\end{align}
Hence, in the original QAML algorithm, the following Ising model Hamiltonian is minimized after transforming the range to $s_i \in \{-1, 1\}$, adding an additional $\lambda$ regularization hyperparameter to penalize nonzero $s_i$~\cite{qml}:
\begin{align}
    H &= \sum_{i=1}^N \left(\lambda - C_i + \frac{1}{2}\sum_{j>i}^N C_{ij}\right)s_i + \frac{1}{4}\sum_{i=1}^N\sum_{j>i}^N C_{ij}s_is_j ,
\end{align}

We observe the following limitations in the QAML algorithm: i) arbitrary linear combinations of weak classifiers $c_i$ are forbidden because the strong classifier $R$ is simply formed by turning weak classifiers $c_i$ on or off; ii) the diversity of the ensemble is limited by the selection of weak classifiers. If the set of weak classifiers can be expanded, more nuanced ensembles with more complex decision boundaries can be learned.

\subsection{Zooming Extension}%
\begin{figure}[ht]
\centering
\includegraphics[width=\columnwidth]{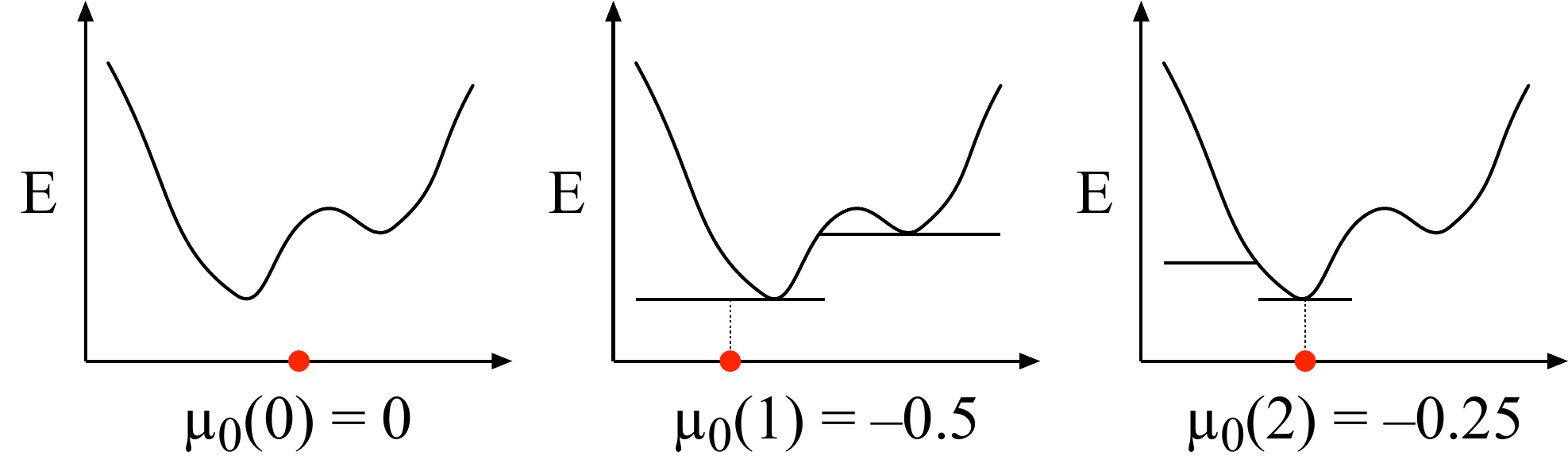}
\caption{\textbf{Zooming extension.} While QAML only performs one anneal, QAML-Z iteratively updates the weight $\mu$ (indicated by the red dot) of a weak classifier (index 0 in the diagram) in the strong classifier ensemble by performing a binary search over the energy surface using spin up/down outcomes.}
\label{fig:zoom}
\end{figure}

By iteratively performing quantum annealing, the binary weights on the weak classifiers can be made continuous, resulting in a stronger classifier. This is achieved by performing a search on the real numbers, effectively zooming in on a region of the energy surface each iteration (Figure~\ref{fig:zoom}). We denote the zooming variant of quantum annealing for machine learning as QAML-Z. Under this reformulation, the weights of the classifiers may be extended from the set $\{0, 1\}$ to the continuous interval $[-1, 1]$, enabling the subtraction of classifiers to reduce cross-correlations between weak classifiers.
\vspace{2mm}

Let each qubit have a mean $\mu_i(t)$ (starting at $\mu_i(0) = 0$ for all $i$) and let the search breadth be $\sigma(t) = b^t$, where $t=0, 1, ..., T-1$ for $T$ iterations and $0 < b < 1$ is a free parameter. Each iteration, the Hamiltonian is centered around the previous mean and the search breadth is narrowed. Receiving spin up or spin down corresponds to shifting the new mean either right or left by a distance given by the search breadth. The weight given to each classifier is thus updated according to the old mean and consequent shift, resulting in a modified Hamiltonian according to the substitution $s_ic_i(\mathbf{x}_\tau) \to \sigma(t) s_ic_i(\mathbf{x}_\tau) + \mu_i(t)c_i(\mathbf{x}_\tau)$. The full expression is: 
%shown in Eq.~\eqref{eq:full}. 
\bes
\begin{align}
    H(t) &= \sum_{i=1}^N\sum_{j>i}^N \sum_{\tau=1}^S \big(\sigma(t)s_ic_i(\mathbf{x}_\tau) + \mu_i(t)c_i(\mathbf{x}_\tau)\big)\notag\\
    &\;\;\;\;\;\;\;\;\;\;\;\;\;\;\;\;\;\;\;\;\;\;\;\times\big(\sigma(t)s_jc_j(\mathbf{x}_\tau) + \mu_j(t)c_j(\mathbf{x}_\tau)\big) \notag\\
    &\;\;\;\;- \sum_{i=1}^N\sum_{\tau=1}^S\big(\sigma(t)s_ic_i(\mathbf{x}_\tau) + \mu_i(t)c_i(\mathbf{x}_\tau)\big)y_\tau \label{eq:full}\\
    &= \sum_{i=1}^N \left(-C_i + \sum_{j=1}^N \mu_j(t)C_{ij} \right)\sigma(t) s_i \notag\\
    &\;\;\;\;+ \sum_{i=1}^N\sum_{j>i}^N C_{ij} \sigma^2(t) s_is_j ,
    \label{eq:H-zoom}
\end{align}
\ees
where Eq.~\eqref{eq:H-zoom} is derived after dropping constants from the Hamiltonian and applying the same $C_i$ and $C_{ij}$ notation as in QAML.
%, we reach . 
This new Hamiltonian may be iteratively optimized for $t=0, 1, \dots, T-1$ to update $\mu_i(t+1) = \mu_i(t) + s_i\sigma(t+1)$, resulting in the new strong classifier $R(\mathbf{x}_\tau) = \sum_{i=1}^N \mu_i(T-1)c_i(\mathbf{x}_\tau)$.

Since the zooming algorithm increases the possibility of overfitting, we propose a two-step randomization procedure to regularize the iterative process.
After each iteration, for each qubit $s_i$ that the energy worsens by the update (i.e., $E[\mu_1(t+1), \dots, \mu_i(t+1), \dots, \mu_N(t+1)] > E[\mu_1(t), \dots, \mu_i(t), \dots \mu_N(t)]$), we apply the flip $s_i \to -s_i$ with monotonically decreasing probability $p_f(t)$, similarly to an annealing schedule in classical simulated annealing. Subsequently, all qubits are uniformly randomly flipped from $s_i$ to $-s_i$ with probability $q_f(t)$ where $q_f(t) < p_f(t)$ for all $t$, akin to a cluster-flip move in variants of simulated annealing. 
%This is both preventing the strong classifier from overfitting as well as pushing it out of local minima in an annealing-inspired procedure. 
This serves to prevent the strong classifier from overfitting as well as to push it out of local minima.
The functions $p_f$ and $q_f$ are specified in the supplementary code.

To take full advantage of these continuous weights, we augment the set of original weak classifiers $h_i(\mathbf{x}_\tau)$ that returns a value in $[-1, 1]$. For each $h_i$, multiple classifiers are generated by shifting the threshold to round to $\pm 1$:
\begin{equation}
    c_{il}(\mathbf{x}_\tau) = \textrm{sgn}(h_i(\mathbf{x}_\tau) + \delta l)/N ,
\end{equation}
where $N$ is the number of classifiers, $l \in \mathbb{Z} : -A \leq l \leq A$ is the offset and $\delta$ is the step size. With a larger set of weak classifiers to ensemble into a strong classifier, a more complex decision boundary may be formed.

% \begin{figure}[ht]
% \centering
% \includegraphics[width=0.45\textwidth]{augmentation.pdf}
% \caption{\textbf{Augmentation of classifiers.} By mapping a continuous classifier $h$ to a larger set of binary classifiers, $c_{il}$, QAML-Z can learn a more complex decision boundary than QAML.}
% \label{fig:augment}
% \end{figure}

With the augmented set of classifiers, the Hamiltonian is now given by:
\begin{align}
    H(t) &= \sum_{l=-A}^A \Bigg[ \sum_{i=1}^N \left(-C_{il} + \sum_{j>i}^N \mu_{jl}(t)C_{ijl} \right)\sigma(t) s_{il} \notag\\
    &\;\;\;\;\;\;\;\;\;\;\;\;\;\;\;+ \sum_{i=1}^N\sum_{j>i}^N C_{ijl} \sigma^2(t) s_{il}s_{jl}\Bigg] ,
\end{align}
where $H(t)$ is iteratively optimized for $t=0, 1, \dots$ to update $\mu_{il}(t+1) = \mu_{il}(t) + s_{il}\sigma(t+1)$. Similarly to before, we have defined:
\begin{align}
    C_{ijl} &= \sum_{\tau=1}^S c_{il}(\mathbf{x}_\tau)c_{jl}(\mathbf{x}_\tau), \\
    C_{il} &= \sum_{\tau=1}^S c_{il}(\mathbf{x}_\tau)y_\tau .
\end{align}

Quantum annealing yields a distribution of excited states, allowing the construction of a stronger classifier than one based purely on ground state results. We take the supremum over the set of excited states' background rejection values for each efficiency in the receiver operating characteristic (ROC) curve according to a validation set of equal size to the training set.

In the experimental demonstration of the algorithm, we set an offset of $A=3$ and a step size of $\delta = 0.0075$. Additionally, we set the zoom parameter $b = \frac{1}{2}$ to perform a binary search over the real numbers. Due to the definition of $\sigma(t)$, the marginal impact of each iteration follows an exponential decay. Thus, QAML-Z was trained for only 8 iterations.

\section{Higgs Boson Classification}%
As an application of QAML-Z, we revisit the Higgs optimization problem, in which kinematic variables describing diphoton processes corresponding either to a Higgs boson decay (signal) or other Standard Model processes (background) are used to identify simulated Higgs bosons~\cite{nature}. The simulation of $H\to\gamma\gamma$ is limited to the main process of gluon fusion, while sub-leading contributions are not included. We evaluated the performance of the QAML-Z algorithm on the programmable D-Wave 2X quantum annealer at the University of Southern California's Information Sciences Institute with 1098 physical qubits~\cite{dwave}. The Ising model is generated from weak classifiers developed from kinematic variables such as transverse momentum, pseudorapidity, and the invariant mass of the diphoton system. However, the QAML-Z algorithm augments this set of weak classifiers with regular offsets of the decision boundary, as described above. In our analysis, we seek primarily to use the Higgs classification problem as a context for providing a clear comparison between QAML-Z and state-of-the-art algorithms in both quantum and classical machine learning. A precise measurement of the Higgs mechanism is beyond the scope of this paper.

\subsection{Quantum Annealing on D-Wave}%
From the 1098 physical qubits of the D-Wave 2X annealer, only 33 fully connected logical qubits are available due to the Chimera graph architecture.  Implementing the exact Ising model with all cross-terms on the scale of the Higgs optimization problem would require hundreds of fully connected qubits, therefore we prune the cross-terms in the Ising Hamiltonian, retaining only the largest 5\% of weights. This reduces the sensitivity to analog errors associated with small weights~\cite{Pearson:2019aa} and also allows a minor embedding operation \cite{embed1, embed2, klymko_adiabatic_2012, Cai:2014nx} in combination with the classical polynomial-time $\texttt{fix\_variables}$ procedure in the D-Wave API to program the problem on the quantum annealer. Each logical qubit is mapped to a chain of physical ferromagnetically coupled qubits on the D-Wave device, where the internal coupling of each chain may be set to prevent thermal excitations and other noise from breaking the chain while still ensuring that the Hamiltonian drives the system dynamics \cite{chains}. The coupling within chains is scaled to the largest coupling in the Hamiltonian and decayed with increasing iteration number. Random errors on the local fields and couplers are reduced by randomizing the encoding by sign flips. Annealing is performed with a 5 $\mu$s anneal time, with minimal variation in performance observed for longer anneal times (not shown). The anneal times were selected to attempt to achieve high performance with the shortest anneal times possible using the D-Wave 2X device, suggesting that future quantum annealers may achieve a wall clock time advantage over simulated annealing if the performance is sustained with lower anneal times.

As in QAML, we use an ensemble of excited states to strengthen the classifier. To select the excited states, we place two criteria: a maximum distance $d$ to the lowest-energy state found (i.e., an excited state must have an energy less than $(1 - d) E_\mathrm{ground}$ for $E_\mathrm{ground} < 0$ or less than $(1 + d) E_\mathrm{ground}$ for $E_\mathrm{ground} > 0$), and a maximum total number of excited states $n_e$ to be selected. To prevent an exponential increase in the tree of excited states generated by the zooming algorithm, we also decay the values of $d$ and $n_e$ by iteration number. The final classifier is then defined by maximizing the area under the ROC curve on a validation set (equivalent to the validation set used for DNN hyperparameter tuning), selecting the best-performing excited states for different false positive rates.

\subsection{Results}%

\begin{figure}[H]
\centering
\includegraphics[width=0.9\columnwidth]{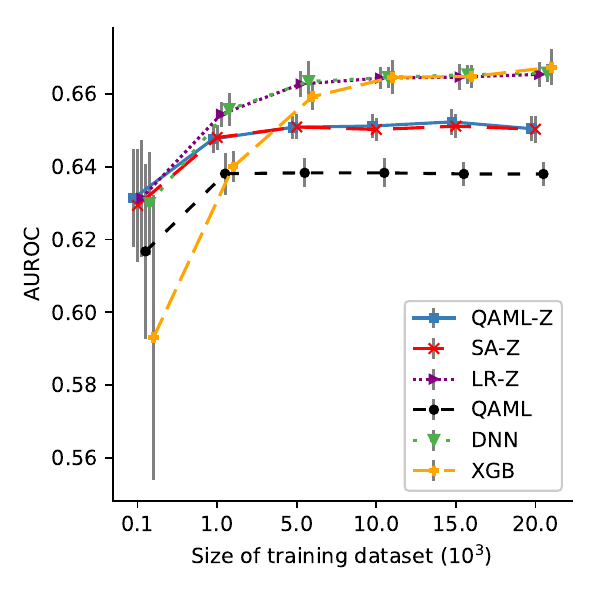}
\caption{\textbf{Area under the ROC curve for the QAML-Z extension, simulated annealing (SA-Z), a logistic regression (LR-Z), the original QAML, a deep neural network (DNN) and XGBoost (XGB)~\cite{10.1145/2939672.2939785} as a function of training set size.} While QAML-Z matches DNN performance at small training set sizes, it decreases the margin between QAML and DNN by 47\% for the largest training sets. Error bars indicate $1\sigma$ error, including both variation over training sets and statistical error estimated by reweighting samples from a Poisson distribution.
}
\label{roc}
\end{figure}

Compared to the QAML algorithm, the area under the receiver operating characteristic curve (AUROC) is significantly improved by QAML-Z on all training set sizes (Figure~\ref{roc}). We select the best-performing classical classifiers (a deep neural network and XGBoost) from the QAML Higgs optimization benchmark, although we optimize additional parameters of the classical algorithms to further improve their performance from Ref.~\cite{nature}. A logistic regression (LR-Z) directly optimizes the mean-squared error of classification over the set of augmented classifiers that QAML-Z is applied to. When compared to classical simulated annealing (SA-Z), QAML-Z performs slightly better (see Figure~\ref{qasa}). 

\begin{figure}[H]
\centering
\includegraphics[width=0.7\columnwidth]{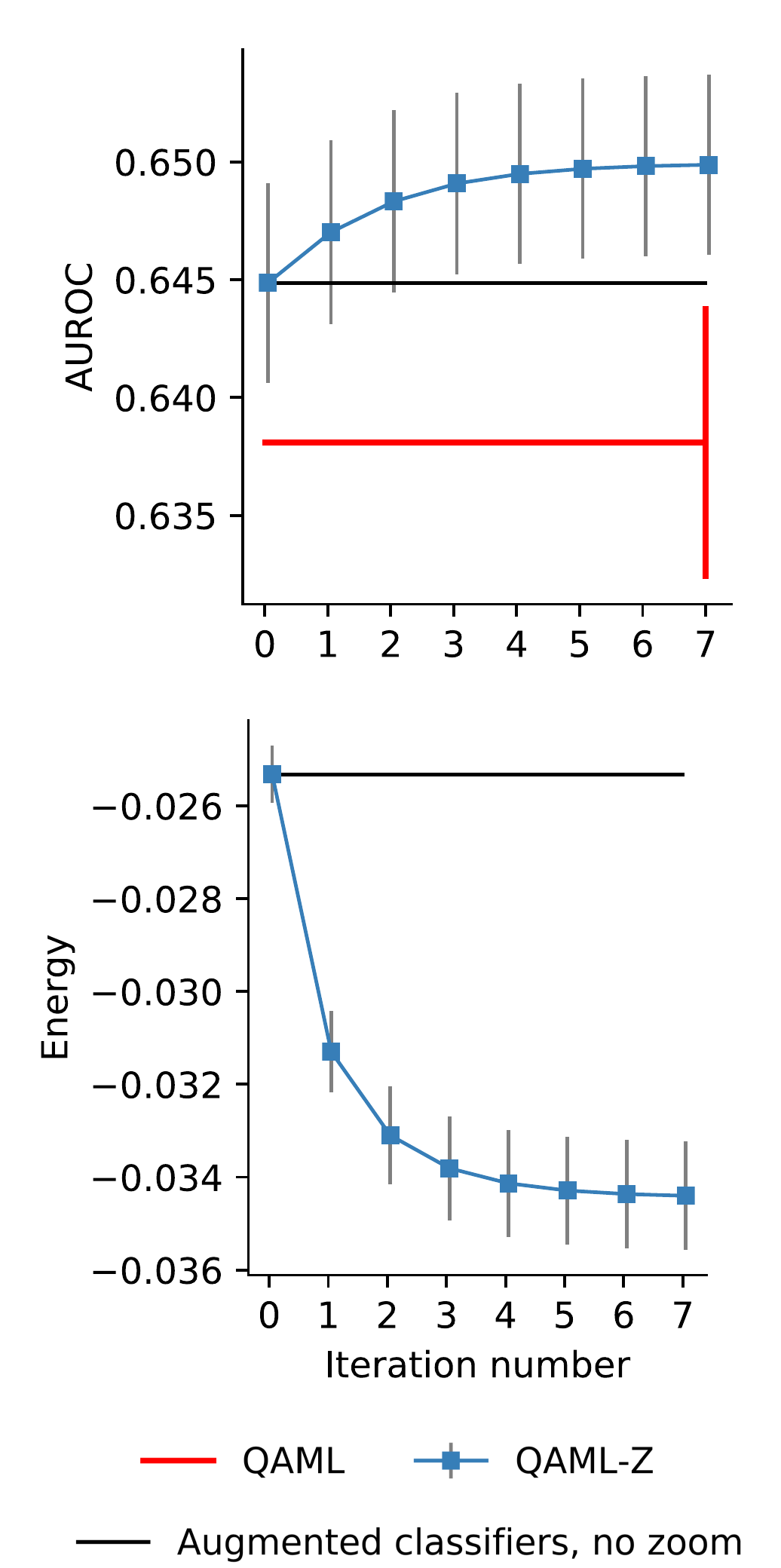}
\caption{\textbf{QAML-Z performance on the test set vs. zooming iteration number (training set size of 1000).} Top: significant improvements by QAML-Z can be separately seen for classifier augmentation (black) and zooming (blue) over the original QAML algorithm (red). Bottom: Ising model energy on the test set improves monotonically, indicating negligible overfitting. Error bars indicate $1\sigma$ error.}
\label{fig:zperform}
\end{figure}

We observe the effectiveness of both the zooming and augmentation aspects of QAML-Z (Figure~\ref{fig:zperform}). The area under the ROC curve illustrates both the impact of classifier augmentation and the impact of zooming, showing advantages in both the classifier augmentation and zooming methodologies. Examining the normalized Ising model energy as a function of iteration number, the zooming algorithm is also shown to monotonically decrease the Hamiltonian energy with additional anneals.

\subsection{Simulated Annealing Benchmark}%
\label{sa}

Given the analogue of quantum annealing to simulated annealing~\cite{kadowaki1998quantum}, we also implement the proposed zooming algorithm in a simulated annealing framework, reporting on simulated annealing with zooming (SA-Z). To attempt to match the improved quantum annealing performance, we also propose simulated annealing with excited states and zooming (SAE-Z), in which the supremum over a set of excited states from simulated annealing is used to improve the area under ROC curve in the same manner as in the quantum algorithm. While a ground state solution minimizes error on the training set, it may overfit to the training data and cause poor generalization on the test set. Hence, the inclusion of excited states --- either thermal noise in simulated annealing or sampled from the quantum annealer --- can improve performance on the test set.

\begin{figure}[H]
\centering
\includegraphics[width=0.9\columnwidth]{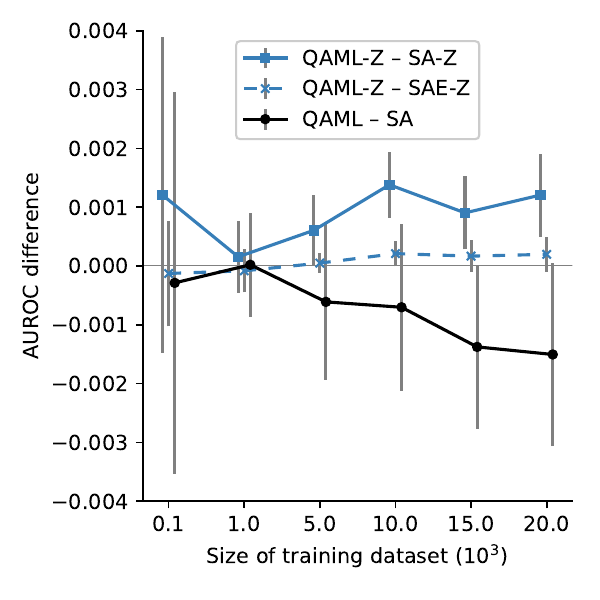}
\caption{\textbf{Comparison of quantum annealing and simulated annealing for the new and original algorithms, measured by area under ROC curve (AUROC).} Although QAML-Z outperforms QAML and SA-Z, the inclusion of excited states in the SAE-Z variant reproduces QAML-Z performance to one standard deviation. Error bars indicate $1\sigma$ error.}
\label{qasa}
\end{figure}

We perform simulated annealing using the Metropolis update rule, flipping a random spin to construct a trial spin vector $\pvec{s}'$ from the spin vector $\vec{s}$.\cite{kirkpatrick_optimization_1983} If the energy $H(\pvec{s}') < H(\vec{s})$, then the new vector $\pvec{s}'$ is accepted with probability 1. However, if $H(\pvec{s}') > H(\vec{s})$, the trial vector is accepted with probability $\exp[-\beta(H(\pvec{s}') - H(\vec{s}))]$. After randomly selecting a spin to flip $N$ times (where $\vec{s}$ has $N$ spins), a sweep has been completed. The inverse temperature $\beta$ is stepped with a linear inverse temperature schedule from $\beta_i = 0.1$ to $\beta_f = 5$ over $W=1000$ sweeps, incrementing the temperature by $\frac{\beta_f - \beta_i}{W}$ after each sweep. This process is repeated 1000 times, and the lowest-energy state is selected in the SA-Z algorithm. Temperature schedules reaching $\beta$ as large as 10 and performing up to 100,000 sweeps per read were found to have no significant impact on the results. To assemble excited states for the SAE-Z benchmark, we perform 5000 sweeps for 5000 reads and select excited states using the same criteria as for quantum annealing.
%, selecting the best $n_e = \{16, 4, 1, 1, \dots\}$ excited states.

\begin{figure}[H]
\centering
\includegraphics[width=0.9\columnwidth]{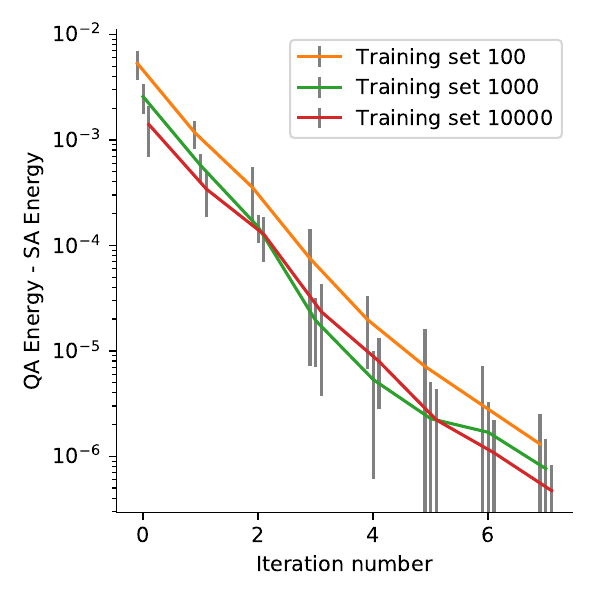}
\caption{\textbf{Difference between the lowest energy of quantum annealing (QA) and simulated annealing (SA).} SA finds a lower minimum energy than QA given an identical initial Hamiltonian. Error bars indicate $1\sigma$ error.}
\label{energy-diff}
\end{figure}

QAML-Z performs better than SA-Z on all training sets, with a statistically significant advantage at larger training set sizes (Figure~\ref{qasa}). This suggests that both simulated and quantum annealing methods found similar ground states at the end of the zooming procedure, although they likely took different paths to the final state due to the fact that SA evolves purely under the classical Hamiltonian, whereas QA evolves under the transverse field as well. When including excited states in simulated annealing, SAE-Z achieves statistically equivalent performance to QAML-Z (Figure~\ref{qasa}), with excited states selected from a validation set improving the generalization ability by reducing overfitting on the training set. A slight discrepancy remains between the two annealing processes, due to the sampling of excited states from distinct distributions of resulting states from simulated and quantum annealing, as well as the analog errors introduced in the implementation on the D-Wave device that are absent in the SA case. However, on the training set, we observe that SA matches or bests QA with regards to minimum observed energy when they are each supplied identical QUBOs generated during the zooming algorithm (Figure~\ref{energy-diff}).

\subsection{Other Classical Benchmarks}%
We provide three additional classical benchmarks to compare QAML-Z performance: an optimized deep neural network (DNN), an optimized XGBoost algorithm (XGB), and a logistic regression (LR-Z). The LR-Z algorithm is a logistic ridge regression on the augmented set of kinematic classifiers used by QAML-Z. The DNN and XGB algorithms are applied to the raw kinematic variables scaled to mean zero and unit standard deviation from the training set and transformed by a principal component analysis to cover 95\% of variance. Just as excited states in QAML-Z are selected by a validation set of equal size to the training set, the hyperparameters of the DNN and XGB are optimized over a similar validation set. While the original DNN benchmark for the Higgs optimization problem~\cite{nature} fixed a given DNN architecture and optimized other hyperparameters, we perform Bayesian optimization over the number of neurons in two hidden layers (2 to 1024 neurons in the first hidden layer and 2 to 4096 neurons in the second hidden layer), L2 regularization ($2^{-12}$ to 0.5) and patience parameter (0.25 to 32) of the DNN. Similarly, we optimize the number of estimators (30 to 10,000), tree depth (1 to 10), learning rate (0.0001 to 0.3), gamma regularization (0 to 15), dataset subsampling (0.2 to 1.0), and feature subsampling (0.3 to 1.0).

\section{Conclusion}%
We find that the QAML-Z extension of quantum annealing over a continuous space of weights on a set of augmented weak classifiers yields strong classifiers that improve the state-of-the-art quantum machine learning algorithm for quantum annealers, which was previously benchmarked in a study of Higgs decay classification~\cite{nature}. Although QAML-Z remains at a disadvantage to a deep neural network (DNN) for sufficiently large datasets, the performance gap between QAML and DNN has been reduced by a factor of two by applying QAML-Z. Moreover, the successful performance at small training set sizes and short 5-$\mu$s anneal times associated with QAML-Z suggests promising applications for online learning on problems that change rapidly, either using a quantum annealer or an FPGA device.

We observe that a logistic regression with LR-Z performs as well as the DNN, which suggests that the augmented set of weak physics-based classifiers is a highly effective method of feature engineering. Although QAML-Z cannot directly minimize least-squared error due to the lack of self-spin terms in the Ising model, it closely matches the performance of LR-Z and DNN at small training set sizes, validating the effectiveness of the quantum annealing approach. Moreover, QAML-Z significantly outperforms an optimized XGBoost algorithm for small training sets, demonstrating competitiveness with state-of-the-art classical machine learning algorithms.

The extent of improvement of QAML-Z over QAML for Higgs decay classification suggests that noisy intermediate-scale quantum devices may be approaching real-world applicability in machine learning despite their limitations. As the fastest annealing time feasible with quantum technology continues to improve, we anticipate that further work on benchmarking wall-clock times of classical and quantum devices will benefit greatly from practically relevant algorithms such as QAML-Z, where performance equal to classical state-of-the-art machine learning has already been demonstrated in certain regimes. More broadly, the favorable results of zooming in on an Ising model to achieve a solution unreachable by discrete optimization provides future direction for quantum annealing applications, potentially extending to quantum machine learning algorithms beyond QAML.

\begin{acknowledgements}
Part of this work was conducted at  ``\textit{iBanks},'' the AI GPU cluster at Caltech. We acknowledge NVIDIA, SuperMicro and the Kavli Foundation for their support of ``\textit{iBanks}.''
This work is partially supported by DOE/HEP QuantISED program grant, Quantum Machine Learning and Quantum Computation Frameworks (QMLQCF) for HEP, award number DE-SC0019227. The work is also supported in part by the AT\&T Foundry Innovation Centers through INQNET, a program for accelerating quantum technologies. The research is based upon work (partially) supported by the Office of the Director of National Intelligence (ODNI), Intelligence Advanced Research Projects Activity (IARPA), via the U.S. Army Research Office contract W911NF-17-C-0050. The views and conclusions contained herein are those of the authors and should not be interpreted as necessarily representing the official policies or endorsements, either expressed or implied, of the ODNI, IARPA, or the U.S. Government. The U.S. Government is authorized to reproduce and distribute reprints for Governmental purposes notwithstanding any copyright annotation thereon.
\end{acknowledgements}

\end{document}